# Distribution Network Congestion Management via Strategic Aggregator Intervention in Local Energy Markets

Ioanna Kalospyrou, Solomon Brown
School of Chemical, Materials and Biological Engineering
University of Sheffield
Sheffield, United Kingdom
ikalospyrou1@sheffield.ac.uk, s.f.brown@sheffield.ac.uk

***Abstract*— High penetration of distributed energy resources increasingly creates congestion in low-voltage distribution networks, while local energy markets (LEMs) optimise community welfare without explicitly internalising network constraints. This paper investigates whether a profit-seeking aggregator embedded within a welfare-oriented LEM can partially internalise distribution-level congestion through market participation. We develop a post-clearing, price-protected intervention in which the aggregator injects additional supply and triggers re-clearing, with network feasibility validated using nonlinear AC power flow subject to a non-deterioration constraint on maximum line loading. The mechanism is benchmarked against Distribution System Operator (DSO)-only corrective control and a hybrid regime with residual DSO action following aggregator intervention. Results on a UK LV feeder show that aggregator participation reduces thermal loading and preserves community welfare relative to DSO-only control, though it does not fully restore compliance under severe stress. The hybrid regime achieves the strongest technical performance while maintaining lower welfare loss. Overall, aggregator intervention remains privately profitable, indicating partial incentive alignment.**

## I. INTRODUCTION

The transition toward low-carbon power systems is accelerating the deployment of distributed energy resources (DERs), including rooftop photovoltaics (PVs), electric vehicles (EVs), and battery energy storage systems (BESS). While these technologies support decarbonisation, their rapid and often uncoordinated uptake increasingly stresses distribution networks [1]. In particular, low-voltage (LV) feeders originally designed for unidirectional flows must now accommodate highly variable, bidirectional injections that push network elements toward their thermal and voltage limits, challenging conventional planning and real-time control [2]. Consequently, distribution-level congestion has emerged as a critical operational concern, motivating research into both conventional control strategies and market-based mechanisms for coordinating distributed flexibility [3], [4].

Local energy markets (LEMs) have been widely proposed as welfare-oriented coordination platforms for distributed flexibility, enabling peer-to-peer trading and improved community-level surplus allocation [5], [6]. In most formulations, market clearing maximises economic efficiency subject to simplified or approximate representations of network constraints, while detailed congestion management remains external to the core clearing mechanism [7].

In parallel, distribution system operators increasingly address congestion and voltage violations through structured flexibility procurement and the development of local flexibility markets (LFMs) [8]. Within LFMs, third-party actors such as aggregators participate as flexibility providers, bundling distributed energy resources and offering controllable demand or storage in response to DSO procurement signals. These markets are explicitly constraint-driven and typically administered by, or closely coordinated with, the DSO. While economically structured, their primary objective is the resolution of network constraints rather than the maximisation of community-level welfare [9]. As a result, economic coordination within LEMs and technical congestion control at the distribution level remain institutionally and operationally separated [7].

A separate stream of research has explored the integration of network feasibility directly into market mechanisms by embedding physical constraints within clearing formulations. This includes centralised, optimisation-based approaches such as congestion-aware clearing, distribution-level locational pricing, and network-constrained optimal power flow models that co-optimise dispatch and network limits in a unified framework [10], [11]. It also encompasses decentralised or distributed market and optimisation framework like peer-to-peer trading mechanisms with AC constraint coordination and distributed optimisation schemes (e.g., ADMM or DLMP-based formulations) that solve for feasibility across agents without full central control [12], [13]. While these approaches represent important progress in coupling economic and physical layers, they are primarily algorithmic solutions for achieving feasibility and consensus rather than models of decentralised market behaviour with embedded strategic, profit-seeking decision-makers.

Taken together, existing approaches provide either welfare-oriented local trading, constraint-driven DSO procurement, or centrally coordinated network-constrained optimisation. However, literature has not yet examined the interaction of profit-seeking participation, welfare clearing, and AC-validated feasibility within a single decentralised architecture.

In particular, it remains unclear whether congestion can be partially internalised through local market coordination itself, rather than being managed exclusively through DSO procurement overlays or fully centralised optimisation. This unresolved interaction between incentives, welfare, and physical feasibility motivates the present study.

We develop a market-embedded congestion management mechanism in which a profit-seeking aggregator participates directly within a welfare-oriented LEM. Rather than operating through a separate flexibility product or DSO procurement channel, the aggregator intervenes post-clearing by submitting controlled additional supply and triggering re-clearing. Network feasibility is validated via nonlinear AC power flow under a non-deterioration constraint. Unlike DLMP or network-constrained OPF approaches, the mechanism preserves uniform-price welfare clearing and enforces physical feasibility sequentially rather than embedding constraints within the optimisation. Performance is benchmarked against stylised DSO corrective control and a hybrid configuration.

This design yields three principal contributions. First, it introduces a profit-seeking aggregator intervention embedded within a LEM via post-clearing redispatch, rather than through a parallel LFM. Second, it implements a non-deterioration-constrained feasibility framework, requiring that maximum line loading does not exceed the baseline outcome, while maintaining uniform-price welfare clearing and avoiding full network-constrained optimisation. Third, it incorporates price-protection logic linked to the pre-intervention clearing price to prevent economic harm to participants. Together, these elements enable systematic evaluation of the interaction between private incentives, community welfare, and AC-validated distribution feasibility within a decentralised market architecture.

## II. METHODOLOGY

### A. System Overview

We consider a radial 0.4 kV LV feeder representative of a UK residential network, with a single slack bus at the MV/LV transformer and several branched laterals supplying domestic customers. The feeder supplies 100 houses arranged in four subfeeders, each comprising five spur branches with five houses connected in series per branch.

A subset of 40 houses forms a Renewable Energy Community (REC) that participates in a Local Energy Market (LEM), while the remaining homes are passive and transact only with the upstream grid. Households may host rooftop PV, EV bidirectional chargers, and in some cases, Battery Energy Storage Systems (BESS), leading to diverse net-load profiles and flexibility across the feeder.

### B. LEM Timeline and Delivery Interval

The study evaluates a single 30-minute delivery interval ($\Delta t$ = 30 min), treated as a snapshot to isolate operational interactions and avoid inter-temporal effects. At T−30 min (gate closure), all community orders are fixed in the LEM order book. The baseline REC order book is cleared, yielding a preliminary price and dispatch schedule under a single-period social-welfare maximisation with uniform pricing.

The resulting injections are applied to the LV network model and evaluated via nonlinear AC power flow (pandapower v2.14.11 [14]) to identify thermal congestion. The aggregator, restricted to seller-side participation (ask orders only), appends its order and the LEM is re-cleared prior to delivery.

The re-cleared schedule is converted into realised injections and applied to the network. If congestion persists and DSO corrective control is enabled, an iterative in-slot corrective loop adjusts realised injections directly. After each corrective action, AC power flow is re-solved until non-deterioration criteria are satisfied.

The final delivered profile, after any DSO correction, constitutes the realised network state for interval [T, T+$\Delta$t]. Settlement is computed by comparing the scheduled market profile with the final delivered profile.

### C. Congestion Relief Sizing for Import-Side Stress

In this work we focus on power flow from the upstream grid into the REC (import side stress). Under the examined scenario set, thermal congestion is the binding network constraint. Relief magnitudes are estimated from overload severity and validated via nonlinear AC power flow.

For each overloaded line $\ell$, with loading $L_\ell > 100\%$, we define overload severity as:

$$a_\ell = \frac{L_\ell - 100}{100} \tag{1}$$

The corresponding active-power estimate (MW) is:

$$\Delta P_\ell^{\mathrm{cong}} = a_\ell \sqrt{3} V_\ell I_{max,\ell} \mathrm{pf} \tag{2}$$

where $V_\ell$ is the nominal line-to-line voltage, $I_{max,\ell}$ is the thermal current limit, and pf = 0.95.

The critical congestion indicator is:

$$P_c^{\mathrm{crit}} = \max_{\ell \in \mathcal{L}} \Delta P_\ell^{\mathrm{cong}} \tag{3}$$

where $\mathcal{L}$ denotes the set of overloaded lines.

The required power relief is therefore defined as

$$P_{req} = P_c^{\mathrm{crit}} \tag{4}$$

### D. Aggregator Intervention Design

The required relief signal $P_{req}$ is communicated to the aggregator following baseline LEM clearing and determination of the preliminary clearing price $p_c$. The aggregator then submits a final ask to the LEM.

The LEM is re-cleared with the augmented order book, yielding an updated clearing price $p_c{}'$ and final dispatch schedule that internalises the aggregator's intervention. As the intervention consists exclusively of additional supply, it shifts the supply curve outward and the clearing price typically decreases or remains unchanged.

The additional supply is delivered via discharge of limited reserved capacity from flexible resources, including EVs and stationary home BESS, contracted to ensure availability without disrupting user activities. This enables the REC to meet excess demand internally rather than increasing upstream imports.

The final aggregator order volume is defined as:

$$P_{order}^{\text{agg}} = \min(P_{req}^{(s)}, P_{flex}, P_{cap}^{LEM}) \quad (5)$$

where $P_{flex}$ is the amount of EV and BESS capacity reserved for potential aggregator activation and $P_{cap}^{LEM}$ is the maximum additional supply that can be cleared without exceeding the pre-intervention clearing price $p_c$.

The effective relief $P_{req}^{(s)}$ is determined under a non-deterioration constraint by scaling the requested relief over a discrete grid $s \in [0,1]$ with step size 0.1. For each candidate scale, the aggregator order is rebuilt, the LEM re-cleared, and the resulting dispatch evaluated via nonlinear AC power flow. The first scale for which maximum line loading does not exceed the baseline (within tolerance) is selected, yielding $P_{req}^{(s)} = sP_{req}$.

The aggregator ask price is selected within the interval $[p_c - \Delta, p_c + \varepsilon]$, where $\varepsilon = 10^{-6}$ ensures a non-degenerate upper bound and $\Delta$ defines the admissible price window. In this study, $\Delta = 4$p/kWh, selected empirically. Candidate prices are drawn from a 10-point uniform grid over this interval, and the price maximising the aggregator's operational profit is selected. Said profit is defined as:

$$\Pi^{\text{agg}} = (p_c' - p_w) P_{cleared}^{\text{agg}} \Delta t \quad (6)$$

where $p_c'$ is the post-intervention clearing price, $p_w$ the wholesale purchase price, $P_{cleared}^{\text{agg}}$ the cleared aggregator volume and $\Delta t$ the interval duration.

### *E. DSO Corrective Action (Import-side Relief)*

Any residual import-driven thermal violations after LEM clearing are mitigated through controlled demand reduction on the critical downstream subtree of the line with highest loading. If the congestion direction is import, the Distribution System Operator (DSO) activates controlled flexibility on eligible loads $\mathcal{J}_{tgt}$; in this study, EV and storage-state loads (i.e., loads with controllable storage state).

The DSO corrective mechanism represents a stylised abstraction of distribution-level flexibility activation for congestion management. In practice, DSOs increasingly procure and activate flexible demand resources (e.g., EV smart charging and storage dispatch) to resolve network constraints prior to curtailment or disconnection. In this model, DSO-only intervention is restricted to demand reduction on eligible flexible loads (EV and storage-state resources) and does not include broader operational measures such as network reconfiguration, tap-changing transformers, or non-targeted curtailment. The formulation therefore captures a constrained flexibility-based corrective logic.

The total controlled demand reduction applied by the DSO is:

$$P_{red}^{(s)} = min\left(P_{req}^{(s)}, \textstyle\sum_{j \in \mathcal{J}_{tgt}} p_j\right) \quad (7)$$

where the term $\sum_{j \in \mathcal{J}_{tgt}} p_j$ represents the total baseline flexible demand available downstream of the critical line.

Demand reduction is then allocated pro rata to baseline load shares across eligible flexible loads:

$$\Delta p_j^{\text{red}} = P_{red}^{(s)} \frac{p_j}{\sum_{k \in \mathcal{J}_{tgt}} p_k} \quad (8)$$

A non-deterioration constraint is then enforced by scaling the requested relief over a discrete grid $s \in [0,1]$, evaluated on a discrete grid with step size 0.1, i.e. $P_{req}^{(s)} = sP_{req}$. For each scaled candidate, nonlinear AC power flow is rerun and only interventions that do not increase maximum line loading relative to baseline are retained. Among feasible candidates, the one minimising post-action maximum loading (followed by lowest overload count and highest retained scale) is selected.

## III. CASE STUDY

Here, we test whether a profit-seeking aggregator within a community LEM can internalise distribution-level congestion and evaluate its performance relative to, and in combination with, DSO corrective control. All cases consider a single 30-minute delivery interval and a wholesale reference price of 8.235 p/kWh, representing the aggregator's procurement cost [15]. Scenarios are designed to induce thermal line loading under UK-specific demand and flexibility conditions while preserving an identical LEM-clearing workflow:

- **Import-biased**: A mildly import-lean operating condition representing a typical winter weekday with moderate demand growth and reduced local generation; baseline seller-side activation case.
- **Evening peak**: A high evening-demand scenario reflecting concentrated residential load, combined with line-rating derating, leading to elevated import congestion.
- **EV super-peak**: An EV-dominated charging event (e.g., post-work clustering), producing the highest demand and V2G penetration; flexibility-rich but stress-intensive.
- **Low-flex**: A stressed import condition with limited BESS and EV-V2G availability, illustrating the operational limits of market-based congestion relief.
- **Compound**: An extreme import day combining high demand, suppressed supply, and tightened network ratings, representing a worst-case congestion regime.

Scenario parameters are summarised in Table 1. The multipliers $s_D$, $s_S$ and $s_{lr}$ scale baseline demand, local generation, and nominal thermal line ratings, respectively. V2G pen. denotes the fraction of REC homes with V2G-capable EVs, while PV+BESS and non-PV+BESS shares indicate the fraction of PV-equipped and non-PV homes with storage. Each scenario is evaluated over five independent runs; results report the mean, as seed variability did not alter qualitative regime rankings.

TABLE I

SCENARIO PARAMETERS AND THEIR VALUES.

| Scenario | $s_D$ | $s_S$ | $s_{lr}$ | V2G pen. | PV+BESS share | Non-PV+BESS share |
|---|---|---|---|---|---|---|
| **Import-biased** | 1.30 | 0.85 | 0.85 | 0.45 | 0.55 | 0.20 |
| **Evening peak** | 1.52 | 0.82 | 0.77 | 0.62 | 0.60 | 0.28 |
| **EV super-peak** | 1.65 | 0.82 | 0.76 | 0.70 | 0.45 | 0.22 |
| **Low-flex** | 1.76 | 0.69 | 0.70 | 0.10 | 0.08 | 0.02 |
| **Compound** | 1.54 | 0.76 | 0.70 | 0.68 | 0.55 | 0.24 |

To isolate the impact of market-based and administrative intervention, we evaluate the following control policies:

- **Aggregator-only** intervention through order submission in LEM,
- **DSO-only** corrective intervention,
- **Hybrid** intervention: aggregator followed by residual DSO action.

## IV. RESULTS

### *A. Congestion Relief Performance*

Across all scenarios, the baseline (Before intervention) maximum line loading exceeds the 100% thermal limit as presented in Fig. 1, confirming that each case induces import-driven congestion. Violations range from approximately 12% to 23% above the limit, with the Evening peak scenario representing the most severe stress condition. The severity of the Evening peak reflects the joint effect of elevated coincident demand and reduced thermal headroom, which amplifies import-driven loading despite moderate flexibility penetration. This indicates that congestion severity is driven not by demand magnitude alone, but by its interaction with available network headroom.

Aggregator-only intervention reduces peak loading in all scenarios, but does not fully restore compliance in high-stress regimes. The largest absolute reduction (approximately 7.5 percentage points) is observed in the Compound scenario. This is consistent with its construction as a high-stress, high-flexibility regime: Compound combines elevated demand and reduced local supply under tightened line ratings, while maintaining high penetrations of dispatchable flexibility. As the aggregator intervenes prior to delivery via LEM re-clearing, additional seller-side EV/BESS discharge is scheduled within the market and directly offsets upstream imports, thereby reducing thermal loading on the critical feeder sections. Notably, Aggregator-only does not increase maximum loading in any scenario, indicating technically well-behaved market intervention under the imposed guardrail.

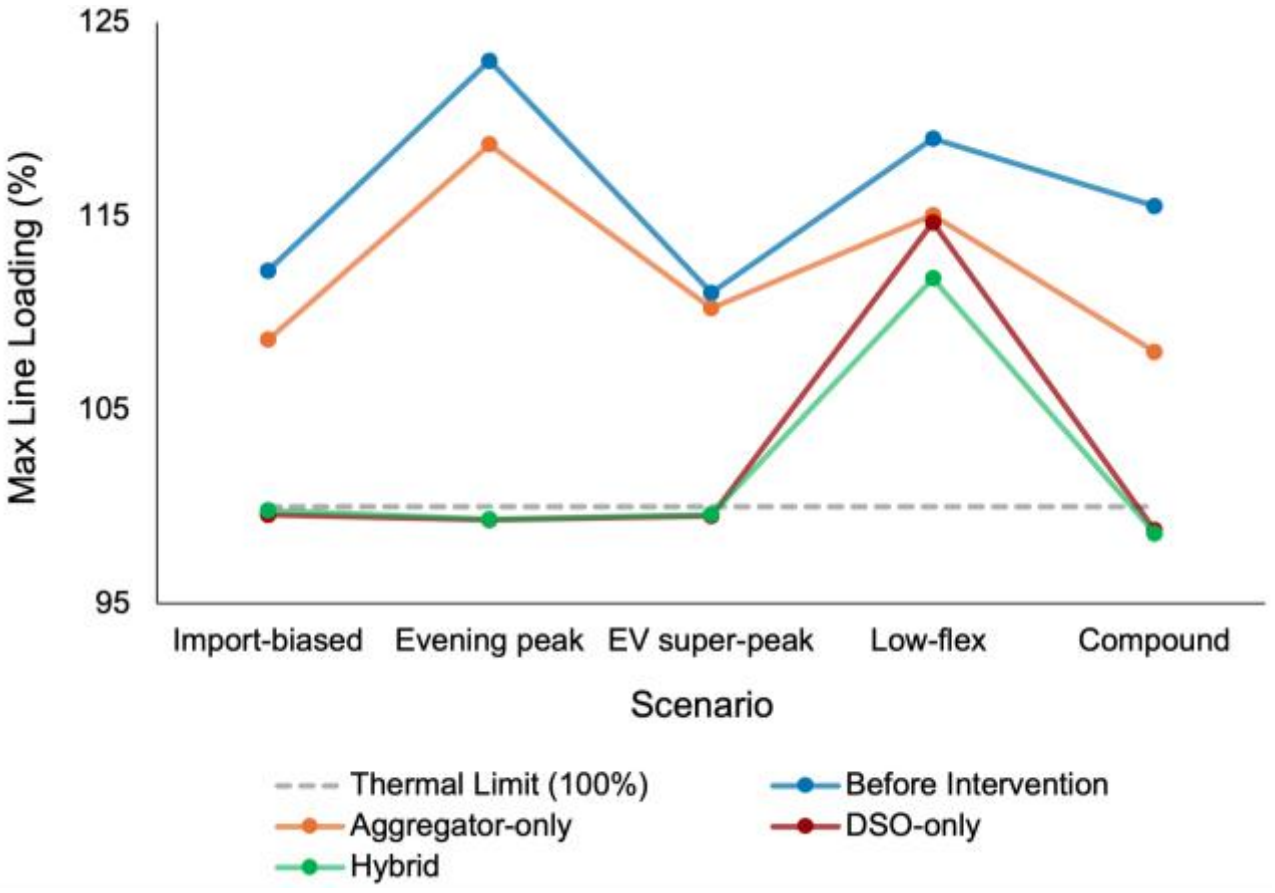

Figure 1: Maximum line loading (%) across scenarios for the baseline case (Before intervention) and under the intervention policies (Aggregator-only, DSO-only, and Hybrid). The dashed line indicates the 100% thermal limit.

DSO-only corrective action restores loading within permissible limits in most scenarios; however, in the Low-flex regime the reduction remains insufficient. In this scenario, high demand coincides with low V2G and BESS penetration, limiting the size of the eligible controllable load set. Under the restricted DSO formulation and the non-deterioration constraint, corrective capacity is limited.

The Hybrid intervention delivers the strongest overall performance. In all scenarios except Low-flex, it restores loading to the thermal limit. In the Low-flex case, it achieves the largest reduction relative to baseline, yet residual violations persist. This outcome indicates a structural flexibility-scarce operating limit, in which both market-based and administrative mechanisms are constrained by available dispatchable capacity rather than by inefficiencies in the intervention design. Taken together, the results reveal two operating regimes: a flexibility-sufficient regime in which combined intervention restores compliance, and a flexibility-scarce regime in which residual congestion reflects structural resource limits.

### *B. Community Welfare Outcomes*

Fig. 2 reports the net community welfare (p per 30-minute interval) computed under each congestion-relief regime (Aggregator-only, Hybrid, and DSO-only). Values are negative across all scenarios because the REC is net-importing in these stressed operating conditions; welfare therefore reflects the net cost of meeting demand through a combination of internal trades and residual external supply and corrective actions.

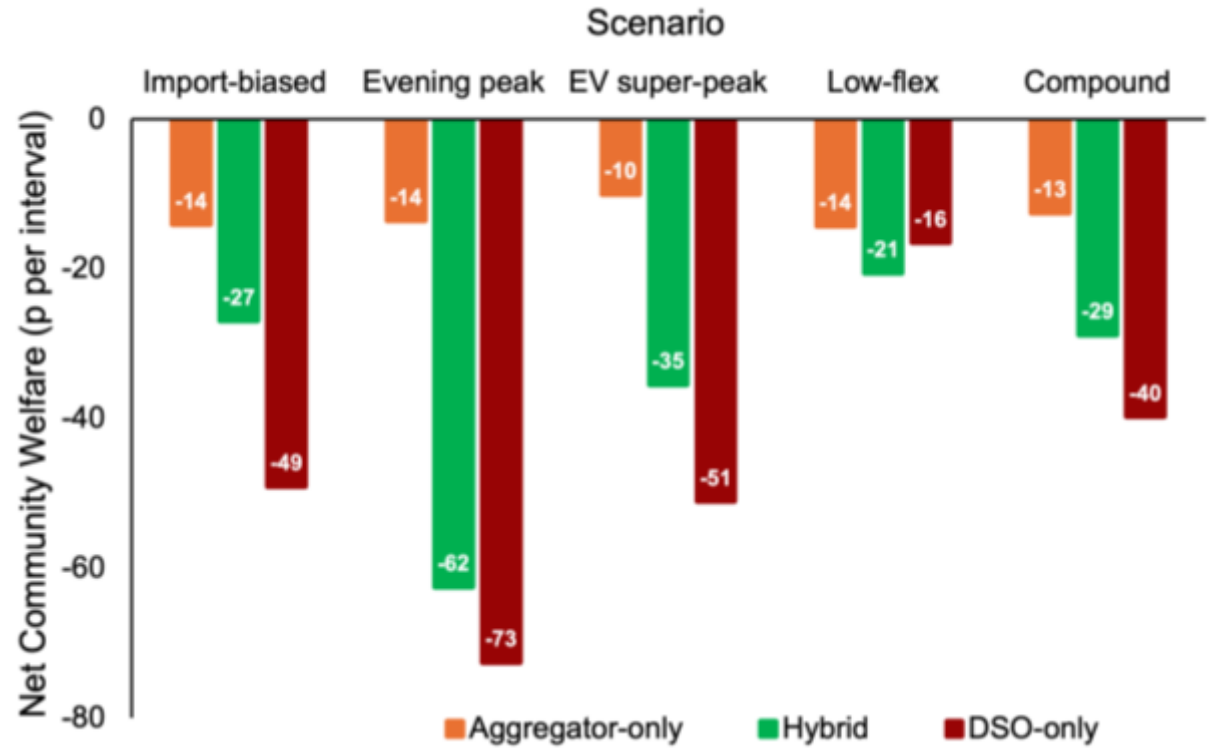

Figure 2: Net community welfare (p per 30min interval, REC total) under Aggregator-only, Hybrid, and DSO-only congestion-relief regimes across scenarios.

Across all scenarios, Aggregator-only yields the least negative welfare, reflecting the efficiency of pre-delivery market re-clearing with seller-side flexibility. In addition, as shown in Fig. 1, Aggregator-only materially reduces maximum line loading in every regime. However, under the highest-stress conditions, market-based intervention alone does not fully restore compliance. These results suggest that while local aggregation can internalise a substantial share of congestion management, complementary corrective mechanisms remain necessary in extreme operating regimes.

DSO-only is the most welfare-costly regime in all scenarios except for Low-flex, consistent with the welfare penalties associated with constraint-driven corrective actions applied outside the market process. The Hybrid regime improves upon DSO-only in these cases, implying that market-based pre-adjustment reduces the extent of residual administrative intervention required to satisfy network limits.

The Low-flex scenario provides a boundary condition. Here, DSO-only yields slightly lower welfare loss than Hybrid (about 5 p per interval), while Aggregator-only remains the least costly but also leaves higher residual congestion than the other interventions. This outcome reflects flexibility scarcity: when residual corrective capacity after market re-clearing is limited, achieving tighter technical compliance requires increasingly welfare-costly intervention.

Joint consideration of Figs 1 and 2 indicates that aggregator participation improves the congestion–welfare trade-off relative to DSO-only control in most scenarios. Although the DSO formulation already represents structured, constraint-driven activation of controllable demand, DSO-only intervention generally incurs the highest community welfare cost (four of five scenarios). Aggregator-only preserves welfare but does not restore thermal compliance under severe stress.

The Hybrid regime combines pre-delivery market re-clearing with residual DSO activation and delivers the strongest technical performance, restoring line loading to within thermal limits in all cases except Low-flex. In welfare terms, Hybrid remains less costly than DSO-only in most scenarios.

Low-flex represents a boundary condition: limited and weakly located flexibility restricts the effectiveness of pre-delivery market re-clearing, such that residual DSO action is still required. As a result, Hybrid can accumulate both market re-dispatch and administrative correction costs, yielding marginally higher welfare loss than DSO-only.

Overall, embedding the aggregator within the LEM shifts part of corrective action from post-delivery administrative control to pre-delivery market scheduling, improving net community welfare in flexibility-sufficient regimes, with Low-flex representing a structural boundary case.

### C. *Aggregator Profitability and Incentive Alignment*

Having evaluated technical performance and community welfare, we now assess whether the proposed aggregator intervention is privately profitable. Using the profit formulation defined in (6), Table II reports mean realised aggregator profit per scenario at the reference wholesale price of 8.235 p/kWh [15]. Values are computed over intervals in which aggregator orders are executed.

Profit is positive in all reported scenarios, indicating that seller-side participation in congestion-relief re-clearing is financially viable when dispatched. Profitability varies across operating regimes and is jointly determined by the price spread between wholesale procurement and post-clearing LEM prices and by the cleared flexibility volume. High-stress conditions can increase potential margins; however, realised profit ultimately depends on the volume actually dispatched through the market.

Importantly, profit arises from pre-delivery market re-dispatch rather than administrative correction. This distinguishes the mechanism from purely regulated congestion management and demonstrates that corrective action can be partially internalised within the market layer. Across all scenarios, aggregator participation generates positive private returns while reducing reliance on welfare-costly DSO intervention, indicating partial alignment between private incentives and congestion mitigation.

From a deployment perspective, this incentive compatibility is critical: absent positive returns, aggregator participation would require regulatory mandate or explicit compensation. While technical effectiveness varies with flexibility availability, the consistent presence of positive profit suggests that distribution-level congestion management can, under suitable conditions, be supported by commercially viable market actors rather than relying exclusively on administrative control.

TABLE II

MEAN AGGREGATOR PROFIT PER SCENARIO FOR A WHOLESALE PRICE OF 8.235 P/KWH.

| Scenario | Aggregator Profit (p per 30min interval) |
|---|---|
| **Import-biased** | 51.90 |
| **Evening peak** | 67.97 |
| **EV super-peak** | 40.69 |
| **Low-flex** | 54.30 |
| **Compound** | 53.51 |

## V. CONCLUSIONS

In this study we investigated whether a profit-seeking aggregator embedded in a community LEM can internalise distribution-level congestion management in a UK-style LV feeder.

Results show that market-based pre-delivery re-clearing reduces thermal loading and preserves community welfare relative to DSO-only corrective control. While Aggregator-only intervention is economically efficient, it does not guarantee full technical compliance under severe stress. The Hybrid regime, combining market scheduling with residual DSO activation, restores loading to thermal limits in flexibility-sufficient conditions and generally delivers the most favourable balance between congestion relief and community welfare. This suggests that local aggregation is a promising mechanism for partially internalising distribution constraints within the market layer. However, it is not universally sufficient on its own: under flexibility-scarce conditions, residual congestion reflects structural resource limits rather than coordination inefficiency.

Importantly, aggregator participation remains privately profitable across all scenarios, indicating partial alignment between private incentives and congestion mitigation.

Future work will extend the model to multi-interval operation with explicit storage dynamics, incorporate bid-side participation for export congestion, and examine scalability effects to assess whether market-only coordination can achieve full compliance in larger or more flexibility-dense communities. The robustness of the aggregator business model under alternative price environments will also be investigated.

### Acknowledgment

Funding from the UK Engineering and Physical Sciences Research Council (EPSRC) and EDF Energy R&D UK Centre Ltd (Grant Reference: EP/Y528808/1) is gratefully acknowledged.

Special thanks to Timothy Hutty for the helpful discussion around aggregator order pricing.

## VI. References

[1] K. K. Mehmood, R. A. Arruda Moura, A. van der Molen, R. Tonkoski, P. Tzscheutschler, P. van der Wielen and P. H. Nguyen, "A review of congestion management methods for power distribution networks: Current practices and future challenges," *Applied Energy,* vol. 407, p. 127342, 15 March 2026.

[2] C. Oerter and N. Neusel-Lange, "LV-grid automation system — A technology review," in *2014 IEEE PES General Meeting | Conference & Exposition*, 2014.

[3] S. Huang, Q. Wu, L. Zhaoxi and A. H. Nielsen, "Review of congestion management methods for distribution networks with high penetration of distributed energy resources," in *IEEE PES Innovative Smart Grid Technologies, Europe*, 2014.

[4] M. Attar, S. Repo, V. Heikkilä and O. Suominen, "Congestion management in distribution grids using local flexibility markets: Investigating influential factors," *International Journal of Electrical Power & Energy Systems,* vol. 167, p. 110601, 2025.

[5] E. Mengelkamp, J. Gärttner, K. Rock, S. Kessler, L. Orsini and C. Weinhardt, "Designing microgrid energy markets: A case study: The Brooklyn Microgrid," *Applied Energy,* vol. 210, pp. 870-880, 15 January 2018.

[6] F. Moret and P. Pierre, "Energy Collectives: A Community and Fairness Based Approach to Future Electricity Markets," *IEEE Transactions on Power Systems,* vol. 34, no. 5, pp. 3994-4004, 2019.

[7] T. Capper, A. Gorbatcheva, M. A. Mustafa, M. Bahloul, J. M. Schwidtal, R. Chitchyan, M. Andoni, V. Robu, M. Montakhabi, I. J. Scott, C. Francis, T. Mbavarira, J. M. Espana and L. Kiesling, "Peer-to-peer, community self-consumption, and transactive energy: A systematic literature review of local energy market models," *Renewable and Sustainable Energy Reviews,* vol. 162, p. 112403, 2022.

[8] A. Scrocca, E. Dacco, D. Andreotti, G. Rancilio, F. Bovera, D. Falabretti and M. Delfanti, "Local flexibility markets in Europe: A critical review of market designs, operational maturity and stakeholder perspectives," *Renewable and Sustainable Energy Reviews,* vol. 226, no. D, p. 116434, 2026.

[9] X. Jin, Q. Wu and H. Jia, "Local flexibility markets: Literature review on concepts, models and clearing methods," *Applied Energy,* vol. 261, p. 114387, 2020.

[10] T. Morstyn, A. Teytelboym, C. Hepburn and M. D. McCulloch, "Integrating P2P Energy Trading With Probabilistic Distribution Locational Marginal Pricing," *IEEE Transactions on Smart Grid,* vol. 11, pp. 3095-3106, 2020.

[11] S. Talari, S. Birk, W. Ketter and T. Schneiders, "Sequential Clearing of Network-Aware Local Energy and Flexibility Markets in Community-Based Grids," *IEEE Transactions on Smart Grid,* vol. 15, pp. 405-417, 2024.

[12] C. Oliveira, M. Simoes, L. Bitencourt, T. Soares and M. A. Matos, "Distributed Network-Constrained P2P Community-Based Market for Distribution Networks," *Energies,* vol. 16, no. 3, p. 1520, 2023.

[13] F. R. Ranjbar, S. N. Ravadanegh and A. Safari, "Transactive energy trading in distribution systems via privacy-preserving distributed coordination," *Applied Energy,* vol. 361, p. 122823, 2024.

[14] Frauhofer IEE and University of Kassel, "pandapower v2.14.11," 2024. [Online]. Available: https://pandapower.readthedocs.io/en/v2.14.11/. [Accessed 18 February 2026].

[15] Ofgem, "Ofgem: Wholesale market indicators," Ofgem, 2025. [Online]. Available: https://www.ofgem.gov.uk/news-and-insight/data/data-portal/wholesale-market-indicators. [Accessed 18 February 2026].